\documentclass[12pt,preprint]{aastex}
\begin{document}
\title{Extreme Magnification Microlensing Event OGLE-2008-BLG-279: Strong Limits on Planetary Companions to the Lens Star}

\author{
J.C.~Yee\altaffilmark{1,2},
A.~Udalski\altaffilmark{3,4},
T.~Sumi\altaffilmark{5,6},
Subo~Dong\altaffilmark{1,2},
S.~Koz{\l}owski\altaffilmark{1,2},
J.C.~Bird\altaffilmark{1,2},
A.~Cole\altaffilmark{7,8},
D.~Higgins\altaffilmark{1,9},
J.~McCormick\altaffilmark{1,10},
B.~Monard\altaffilmark{1,11},
D.~Polishook\altaffilmark{1,12},
A.~Shporer\altaffilmark{1,12},
O.~Spector\altaffilmark{1,12},
\\
and\\
M.\,K.~Szyma{\'n}ski\altaffilmark{4},
M.~Kubiak\altaffilmark{4},
G.~Pietrzy{\'n}ski\altaffilmark{4,13},
I.~Soszy{\'n}ski\altaffilmark{4},
O.~Szewczyk\altaffilmark{13},
K.~Ulaczyk\altaffilmark{4},
{\L}.~Wyrzykowski\altaffilmark{14,4},
R.~Poleski\altaffilmark{4}\\
(The OGLE Collaboration),\\
and\\
W.~Allen\altaffilmark{15},
M.~Bos\altaffilmark{16},
G.W.~Christie\altaffilmark{17},
D.L.~DePoy\altaffilmark{18},
J.D.~Eastman\altaffilmark{2},
B.S.~Gaudi\altaffilmark{2},
A.~Gould\altaffilmark{2,19},
C.~Han\altaffilmark{20},
S.~Kaspi\altaffilmark{12},
C.-U.~Lee\altaffilmark{21},
F.~Mallia\altaffilmark{22},
A.~Maury\altaffilmark{22},
D.~Maoz\altaffilmark{12},
T.~Natusch\altaffilmark{23},
B.-G.~Park\altaffilmark{21},
R.W.~Pogge\altaffilmark{2},
R.~Santallo\altaffilmark{24}\\
(The $\mu$FUN Collaboration),\\
and\\
F.~Abe\altaffilmark{6},
I.A.~Bond\altaffilmark{25},
A.~Fukui \altaffilmark{6},
K.~Furusawa\altaffilmark{6},
J.B.~Hearnshaw\altaffilmark{26},
S.~Hosaka\altaffilmark{6},
Y.~Itow\altaffilmark{6},
K.~Kamiya\altaffilmark{6},
A.V.~Korpela\altaffilmark{27},
P.M.~Kilmartin\altaffilmark{28},
W.~Lin\altaffilmark{25},
C.H.~Ling\altaffilmark{25},
S.~Makita\altaffilmark{6},
K.~Masuda\altaffilmark{6},
Y.~Matsubara\altaffilmark{6},
N.~Miyake\altaffilmark{6},
Y.~Muraki\altaffilmark{29},
M.~Nagaya\altaffilmark{6},
K.~Nishimoto\altaffilmark{6},
K.~Ohnishi\altaffilmark{30},
Y.C.~Perrott\altaffilmark{31},
N.J.~Rattenbury\altaffilmark{31},
T.~Sako\altaffilmark{6},
To.~Saito\altaffilmark{32},
L.~Skuljan\altaffilmark{25},
D.J.~Sullivan\altaffilmark{27},
W.L.~Sweatman\altaffilmark{25},
P.J.~Tristram\altaffilmark{28},
P.C.M.~Yock\altaffilmark{31}\\
({The MOA Collaboration}),\\
and\\
M.D.~Albrow\altaffilmark{26}, 
V.~Batista\altaffilmark{19}, 
P.~Fouqu\'e\altaffilmark{33},
J.-P.~Beaulieu\altaffilmark{19,34}, 
D.P.~Bennett\altaffilmark{35}, 
A.~Cassan\altaffilmark{36}, 
J.~Comparat\altaffilmark{19},
C.~Coutures\altaffilmark{19},
S.~Dieters\altaffilmark{19}, 
J.~Greenhill\altaffilmark{8},
K.~Horne\altaffilmark{37}, 
N.~Kains\altaffilmark{37},
D.~Kubas\altaffilmark{37}, 
R.~Martin\altaffilmark{38}, 
J.~Menzies\altaffilmark{39},
J.~Wambsganss\altaffilmark{36},
A.~Williams\altaffilmark{38}, 
M.~Zub\altaffilmark{36}\\
({The PLANET Collaboration})\\
}

\altaffiltext{1}
{Microlensing Follow Up Network ($\mu$FUN)}
\altaffiltext{2}
{Department of Astronomy, Ohio State University,
140 W.\ 18th Ave., Columbus, OH 43210, USA; 
dong,gaudi,gould,jyee,pogge@astronomy.ohio-state.edu}
\altaffiltext{3}
{Optical Gravitational Lens Experiment (OGLE)}
\altaffiltext{4} {Warsaw University Observatory, Al.~Ujazdowskie~4, 00-478~Warszawa, Poland; udalski,msz,mk,pietrzyn,soszynsk,kulaczyk,rpolesk@astrouw.edu.pl}
\altaffiltext{5}{Microlensing Observations in Astrophysics (MOA)}
\altaffiltext{6}{Solar-Terrestrial Environment Laboratory,
    Nagoya University, Nagoya 464-8601, Japan; sumi@stelab.nagoya-u.ac.jp}
\altaffiltext{7}{Probing Lensing Anomalies NETwork (PLANET)}
\altaffiltext{8}{School of Mathematics and Physics, University of Tasmania,
Private Bag 37, Hobart, Tasmania 7001, Australia}
\altaffiltext{9}
{Hunters Hill Observatory, Canberra, Australia; higginsdj@bigpond.com}
\altaffiltext{10}
{Farm Cove Observatory, Centre for Backyard Astrophysics,
Pakuranga, Auckland, New Zealand; farmcoveobs@xtra.co.nz}
\altaffiltext{11}
{Bronberg Observatory, Centre for Backyard Astrophysics, Pretoria, South
Africa; lagmonar@nmisa.org}
\altaffiltext{12}{School of Physics and Astronomy and Wise Observatory, Tel-Aviv University, Tel-Aviv 69978, Israel}
\altaffiltext{13} {Universidad de Concepci{\'o}n, Departamento de Fisica, Casilla 160--C, Concepci{\'o}n, Chile; szewczyk@astro-udec.cl}
\altaffiltext{14} {Institute of Astronomy, University of Cambridge, Madingley Road, Cambridge CB3 0HA, UK; wyrzykow@ast.cam.ac.uk}
\altaffiltext{15}
{Vintage Lane Observatory, Blenheim, New Zealand; whallen@xtra.co.nz}
\altaffiltext{16}
{Molehill Astronomical Observatory, Auckland, New Zealand; molehill@ihug.co.nz}
\altaffiltext{17}{Auckland Observatory, Auckland, New Zealand; gwchristie@christie.org.nz}
\altaffiltext{18}
{Dept.\ of Physics, Texas A\&M University, College Station, TX, USA; 
depoy@physics.tamu.edu}
\altaffiltext{19}
{Institut d'Astrophysique de Paris UMR7095, 98bis Boulevard Arago, 75014,
Paris, France; beaulieu,coutures@iap.fr}
\altaffiltext{20}
{Department of Physics, Institute for Basic Science Research,
Chungbuk National University, Chongju 361-763, Korea;
cheongho@astroph.chungbuk.ac.kr}
\altaffiltext{21}
{Korea Astronomy and
Space Science Institute, Daejon 305-348, Korea; leecu,bgpark@kasi.re.kr}
\altaffiltext{22}
{Campo Catino Austral Observatory, San Pedro de Atacama, Chile; francomallia@campocatinobservatory.org}
\altaffiltext{23}
{AUT University, Auckland, New Zealand; tim.natusch@aut.ac.nz}
\altaffiltext{24}
{Southern Stars Observatory, Faaa, Tahiti, French Polynesia; santallo@southernstars-observatory.org}
\altaffiltext{25}
{Institute of Information and Mathematical Sciences, Massey University,
Auckland, New Zealand; i.a.bond@massey.ac.nz}
\altaffiltext{26}{Department of Physics and Astronomy,
    University of Canterbury, Private Bag 4800,
   Christchurch, New Zealand}
\altaffiltext{27}{School of Chemical and Physical Sciences,
     Victoria University, Wellington, New Zealand}
\altaffiltext{28}{Mt. John Observatory, P.O. Box 56,
      Lake Tekapo 8770, New Zealand}
\altaffiltext{29}{Konan University, Kobe, Japan}
\altaffiltext{30}{Nagano National College of Technology,
      Nagano 381-8550, Japan}
\altaffiltext{31}{Department of Physics, University of Auckland,
     Auckland, New Zealand}
\altaffiltext{32}{Tokyo Metropolitan College of Aeronautics,
      Tokyo 116-8523, Japan}
\altaffiltext{33}{LATT, Universit\'{e} de Toulouse, CNRS, 14 avenue Edouard Belin,
31400 Toulouse, France}
\altaffiltext{34}{Department of Physics and Astronomy, University College London,
Gower Street, London, WC1E 6BT, United Kingdom}
\altaffiltext{35}{Department of Physics,
    University of Notre Dame, IN 46556, USA; bennett@nd.edu}
\altaffiltext{36}{Astronomisches Rechen-Institut (ARI), Zentrum f\"{u}r Astronomieder Universit\"{a}t Heidelberg (ZAH), M\"{o}nchhofstrasse 12­-14, 69120 Heidelberg, Germany}
\altaffiltext{37}{Scottish Universities Physics Alliance, School of Physics \&
Astronomy, University of St Andrews, North Haugh, St Andrews,
KY16~9SS, United Kingdom}
\altaffiltext{38}{Perth Observatory, Walnut Road, Bickley, Perth 6076, Australia}
\altaffiltext{39}{South African Astronomical Observatory, P.O. Box 9 Observatory
7935, South Africa}

\newcommand{\bdv}[1]{\mbox{\boldmath$#1$}}
\def\bpi{{\bdv{\pi}}}

\begin{abstract}
We analyze the extreme high-magnification microlensing event OGLE-2008-BLG-279, which peaked at a maximum magnification of $A \sim 1600$ on 30 May 2008.  The peak of this event exhibits both finite-source effects and terrestrial parallax, from which we determine the mass of the lens, $M_l=0.64 \pm 0.10 M_{\odot}$, and its distance, $D_l = 4.0 \pm 0.6\,\mathrm{kpc}$. We rule out Jupiter-mass planetary companions to the lens star for projected separations in the range 0.5-20 AU. More generally, we find that this event was sensitive to planets with masses as small as $0.2~M_\oplus \simeq 2~M_{\rm {Mars}}$ with projected separations near the Einstein ring ($\sim 3~{\rm AU}$).

\end{abstract}


\section{Introduction}
A complete census of planets beyond the snow line will be crucial for testing the currently favored core-accretion theory of planet formation since that is the region where this model predicts that giant planets form. For example, \citet{Ida04} find that gas giant planets around solar-type stars preferentially form in the region between the snow line at 2.7 AU and $\sim 10\,$ AU. While radial velocity and transit searches account for most of the more than 300 planets known to date, microlensing has the ability to probe a different region of parameter space that reaches far beyond the snow line and down to Earth-mass planets. Microlensing is most sensitive to planets near the Einstein ring radius, which \citet{Gould92} showed lies just outside the snow line:
\begin{equation}
r_{\rm{E}} \simeq 4\left(\frac{M_l}{M_{\odot}}\right)^{1/2} \mathrm{\,AU},
\end{equation}
for reasonable assumptions. This sensitivity to planets beyond the snow line is demonstrated by the eight published planets found by microlensing, which range in mass from super-Earths to Jupiters and more massive objects \citep{Bond04, Udalski05, Beaulieu06, Gould06, Bennett08, Dong09, Gaudi08}.

In high magnification microlensing events ($A \gtrsim 100$), the images finely probe the full angular extent of the Einstein ring, making these events particularly sensitive to planets over a wide range of separations \citep{Griest98}. Additionally, because the time of maximum sensitivity to planets (the peak of the event) can be determined in advance, intensive observations can be planned resulting in improved coverage of the event, particularly given limited resources. Even when a planet is not detected, the extreme sensitivity of such an event can be used to put broad constraints on planetary companions.

High magnification events are also useful because it is more likely that secondary effects such as the finite-source effect and terrestrial parallax can be measured \citep{Gould97}. These effects can be used to break several microlensing degeneracies and allow a measurement of the mass of the lens and its distance. This allows us to determine a true mass of a planet rather than the planet/star mass ratio and a true projected separation rather than a relative one. Thus, in addition to being more sensitive to planets, high magnification events allow us to make more specific inferences about the nature of the system.

Previous work has empirically demonstrated the sensitivity of high
magnification events to giant planets by analyzing observed events
without detected planets and explicitly computing the detection
sensitivity of these events to planetary companions. The first high
magnification event to be analyzed in such a way was MACHO 1998-BLG-35
\citep{Rhie00}. \citet{Rhie00} found that planets with a Jupiter-mass
ratio ($q=10^{-3}$) were excluded for projected separations in units
of the Einstein ring radius of $d=$0.37--2.70.  Since then, many other
authors have analyzed the planet detection sensitivity of individual
high-magnification events \citep{Bond02,Gaudi02,Yoo04,Abe04,Dong06,Batista09}.
In particular, prior to the work presented here, the most sensitive
event with the broadest constraints on planetary companions was MOA
2003-BLG-32, which reached a magnification of 520
\citep{Abe04}. \citet{Dong06} found that this
event had sensitivity to giant planets out to $d\lesssim 4$\footnote{\citet{Dong06} also analyzed the event
OGLE-2004-BLG-343, which reached a peak magnification of $A\sim 3000$. Although this is the highest magnification event
analyzed for planets, sparse observational coverage over the peak
greatly reduced its sensitivity.}.

This paper presents the analysis of OGLE-2008-BLG-279, which reached a
magnification of $A \sim 1600$ and was well-covered over the peak,
making it extremely sensitive to planetary companions.  In fact, as we
will show, this event has the greatest sensitivity to planetary
companions of any event yet analyzed, and we can exclude planets over
a wide range of separations and masses.  Furthermore, this event
exhibited finite-source effects and terrestrial parallax, allowing a
measurement of the mass and distance to the lens.  This allows us to
place constraints on planets in terms of their mass and projected
separation in physical units.  We begin by
describing the data collection and alert process in
$\S$\ref{sec:data}. In $\S$\ref{sec:fits} we describe our fits to the
light curve and the source parameters. We then go on to discuss the
blended light and the shear contributed by a nearby star in
$\S$\ref{sec:blend}. Finally, we place limits on planetary companions
in $\S$\ref{sec:planets}. We conclude in $\S$\ref{sec:conclusions}.

\section{Data Collection}
\label{sec:data}
\begin{figure}
\includegraphics[width=5.5in]{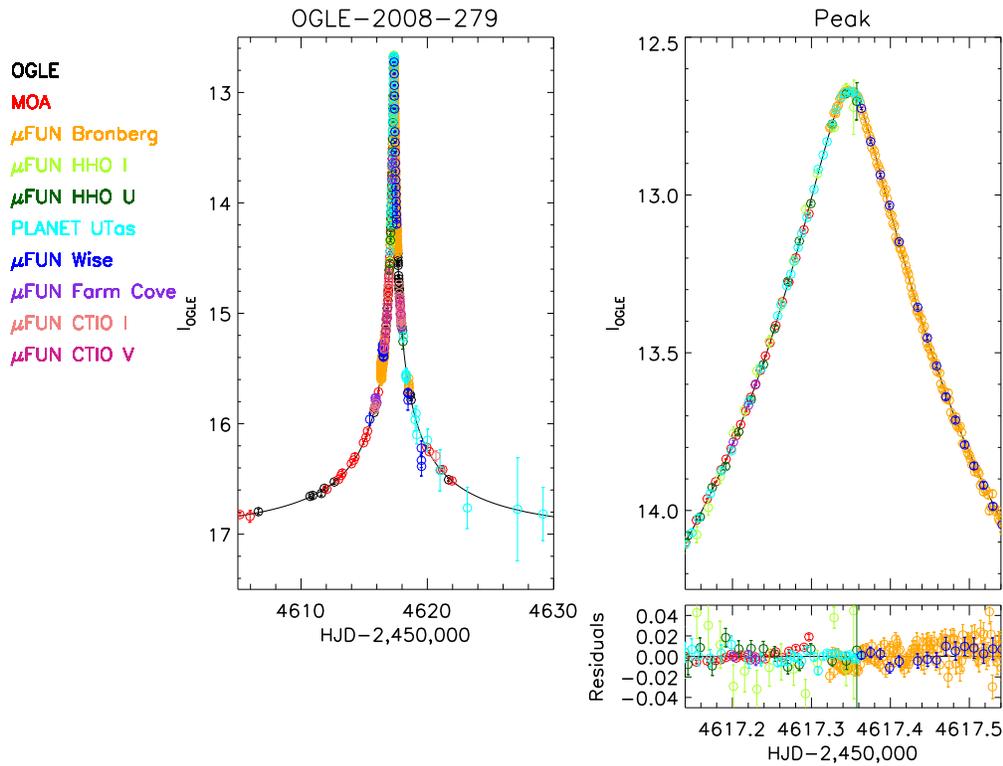}
\caption{Light curve of OGLE-2008-BLG-279 near its peak. The left panel shows the entire event, while the right panel shows a close-up of the peak with residuals from the point-lens model including finite-source effects. The black solid line shows this best-fit model. For clarity, the data have been binned and rescaled to the OGLE flux. \label{fig:lc}}
\end{figure}

On 2008 May 13 (HJD$'$ $\equiv$ HJD - 2,450,000 = 4600.3604), the OGLE collaboration announced the discovery of a new microlensing event candidate OGLE-2008-BLG-279 at RA=17$^{\rm h}$58$^{\rm m}$36.\hskip-2pt$^{\rm s}$17 Dec=-30$^{\circ}22'08.\hskip-2pt''4$ (J2000.0). This event was independently announced by the MOA collaboration on 2008 May 26 as MOA-2008-BLG-225. Based on the available OGLE and MOA data, $\mu$FUN began observations of this event on 2008 May 27 from the CTIO SMARTS 1.3m in Chile, acquiring observations in both the $V$ and $I$ bands, and the next day identified it as likely to reach very high magnification two days hence. This event was monitored intensively over the peak by MOA, the PLANET collaboration, and many $\mu$FUN observatories. Specifically, the $\mu$FUN observatories Bronberg, Hunters Hill, Farm Cove, and Wise obtained data over the peak of this event  (see Fig. \ref{fig:lc}). OGLE-2008-BLG-279 peaked on 2008 May 30 at HJD$'$ = 4617.3481 with a magnification $A \sim 1600$.

Because there were so many data sets, this analysis focuses on the $\mu$FUN data from observatories that covered the peak of the event ($\mu$FUN Bronberg (South Africa), Hunters Hill (Australia), Farm Cove (New Zealand), and Wise (Israel)) and PLANET Canopus (Australia) combined with the data from OGLE and MOA which cover both peak and baseline. We used the data from CTIO to measure the colors of the event but not in other analyses. Early fits of the data indicated that the Bronberg data from HJD$'$4617.0-4617.32 suffer from systematic residuals that are more severe than those seen in any of the other data, so these data were excluded from subsequent analysis.

The data were all reduced using difference imaging analysis \citep[DIA;][]{Wozniak00} with the exception of the CTIO data which were reduced using the DoPHOT package \citep{Schechter93}. The uncertainties in all the data sets were normalized so that the $\chi^2$/degree-of-freedom $\sim$ 1, and we removed $>3\sigma$ outliers whose deviations were not confirmed by near simultaneous data from other observatories. The normalization factors for each observatory are as follows: OGLE(1.8), MOA(1.0), $\mu$FUN Bronberg(1.4), $\mu$FUN Hunters Hill I(2.7) and U(1.5), $\mu$FUN Farm Cove(2.1), $\mu$FUN Wise(3.8), PLANET Canopus(4.6), and $\mu$FUN CTIO I(1.4) and V(2.0).

\section{Point-Lens Analysis}
\label{sec:fits}
\begin{deluxetable}{cccccrlrrlrr}
\rotate
\tablewidth{0pt}
\tablecaption{Light Curve Fits \label{tab:params}}
\tablehead{\multicolumn{4}{c}{Effects}&\colhead{}&\multicolumn{7}{c}{Fit Parameters}\\ \cline{1-4}\cline{7-12}
\colhead{Finite-}&\colhead{Orbital}&\colhead{Terrestrial}&\colhead{}&\colhead{}&\colhead{}&\colhead{$t_0-4617.34$ }&\colhead{$u_0$}&\colhead{$t_{\mathrm{E}}$}&\colhead{$\rho_{\star}$}&\colhead{$\pi_{\rm{E},E}$}&\colhead{$\pi_{\rm{E},N}$}\\
\colhead{Source}&\colhead{Parallax}&\colhead{Parallax}&\colhead{$-u_0$}&\colhead{} &\colhead{$-\Delta\chi^2$}&\colhead{(days)}&\colhead{$(\theta_{\rm E})$}&\colhead{(days)}&\colhead{$(\theta_{\rm E})$}&\colhead{}&\colhead{}\\
\colhead{[1]}&\colhead{[2]}&\colhead{[3]}&\colhead{[4]}&\colhead{}&\colhead{[5]}&\colhead{[6]}&\colhead{[7]}&\colhead{[8]}&\colhead{[9]}&\colhead{[10]}&\colhead{[11]}
}

\startdata
\checkmark & & & & &0.00 &0.00783(7) & 6.4(5)$\times10^{-4}$ & 111.(9)\phn & 6.6(6)$\times10^{4}$ & \nodata & \nodata\\
\checkmark &\checkmark &\checkmark & & & 164.50&0.00787(8) & 6.6(5)$\times10^{-4}$ & 106.(9)\phn &  6.8(6)$\times10^{4}$ & -0.15(2) &  0.02(2)\\
\checkmark &\checkmark &\checkmark &\checkmark & & 127.97&0.0081(1) & -6.4(6)$\times10^{-4}$ & 109.(9)\phn & 6.7(6)$\times10^{4}$ & 0.11(2) & 0.09(2)\\
\checkmark &\checkmark & & & & 15.52& 0.00784(8) & 8.(1)\phn$\times10^{-4}$ & \phn84.(12) & 9.(1)\phn$\times10^{4}$ & 1.5(4)\phn & -0.3(2)\phn\\
\checkmark &\checkmark & & \checkmark & & 15.51&0.00786(8)& -8.(1)\phn$\times10^{-4}$ & \phn84.(12) & 9.(1)\phn$\times10^{4}$ & 1.5(4)\phn & -0.3(2)\phn\\
\checkmark & &\checkmark & & & 166.40&0.00787(8)& 6.9(6)$\times10^{-4}$ & 101.(8)\phn & 7.2(6)$\times10^{4}$ & -0.16(2) & 0.03(2)\\
\checkmark & &\checkmark &\checkmark & & 129.59 &0.0081(1)& -6.9(5)$\times10^{-4}$ & 102.(8)\phn & 7.1(6)$\times10^{4}$ & 0.11(2) & 0.11(3)\\
\enddata

\tablecomments{The first 4 columns indicate which effects were included in the point-lens fit. The $\Delta\chi^2$ improvement for each fit (col. 5) is given relative to the best-fit including finite-source effects but without parallax. There are 5731 data points in the fit light curve. The numbers in parentheses indicate the uncertainty in the final digit or digits of the fit parameters.}
\end{deluxetable}

The data for OGLE-2008-BLG-279 appear to be consistent with a very-high magnification, $A=1570\pm120$, single lens microlensing event. We therefore begin by fitting the data with a point-lens model and then go on to place limits on planetary companions in $\S$\ref{sec:planets}. In this section, we describe our fits to the data and address the second-order, finite-source and parallax effects on the light curve. 

\subsection{Angular Einstein Ring Radius}
\label{sec:source}
\begin{figure}
\includegraphics[width=5.5in]{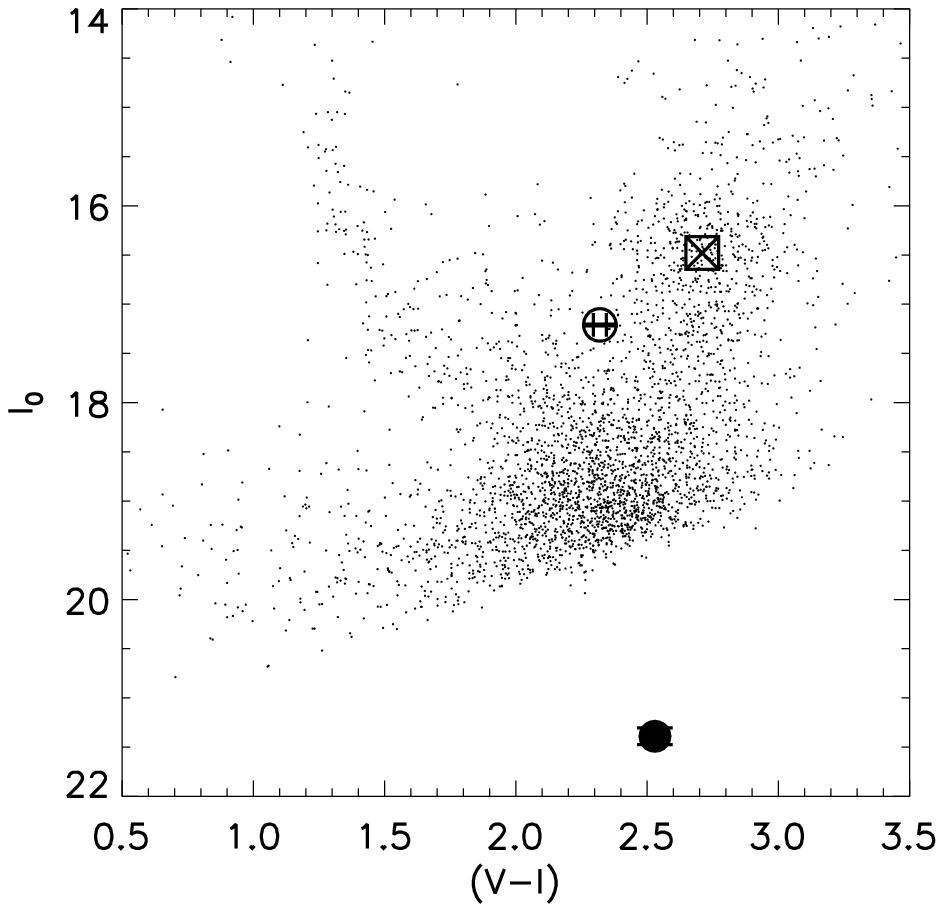}
\caption{Calibrated Color-Magnitude Diagram (CMD) constructed from the CTIO and OGLE data. The square indicates the centroid of the red clump, the open circle shows the blended light, and the solid circle indicates the source. The small black points are field stars. The error bars are shown but are smaller than the size of the points. \label{fig:cmd}}
\end{figure}
From the $V$- and $I$-band images taken with CTIO both during the peak and after the event, we construct a CMD of the event (Fig. \ref{fig:cmd}). We calibrate this CMD using stars that are also in the calibrated OGLE-III field. For the source, we measure $[I, (V-I)] = [21.39 \pm 0.09, 2.53 \pm 0.01]$. If we assume that the source is in the bulge and thus behind the same amount of dust as the clump, we can compute the dereddened color and magnitude. We measure the color and magnitude of the clump: $[I, (V-I)]_{\rm{cl}}=[16.48, 2.71]$. The absolute color and magnitude of the clump are $[M_I, (V-I)_0]_{\rm{cl}} = [-0.20, 1.05]$, which at a distance of 8.0 kpc would appear to be $[I, (V-I)]_{0,\rm{cl}} = [14.32, 1.05]$. We find $A_I = I_{\rm{cl}} - I_{0,\rm{cl}} = 16.48 - 14.32 = 2.16$ and $E(V-I) = (V-I)_{\rm{cl}} - (V-I)_{0,\rm{cl}} = 1.66$. We then calculate the dereddened color and magnitude of the source to be $[I, (V-I)]_0 = [19.23, 0.87]$.

The angular Einstein ring radius can be determined by combining information from the light curve and the color-magnitude diagram (CMD). Finite source effects in the light curve enable us to determine the ratio of the source size, $\theta_{\star}$, to the Einstein radius, $\theta_{\rm{E}}$:
\begin{equation}
\label{eqn:rho}
\rho_{\star} = \theta_{\star}/\theta_{{\rm E}}.
\end{equation}
We can then estimate $\theta_{\star}$ from the color and magnitude of the source measured from the CMD, and solve for $\theta_{\rm{E}}$.

\subsubsection{Finite-Source Effects}

\begin{figure}
\includegraphics[width=5.5in]{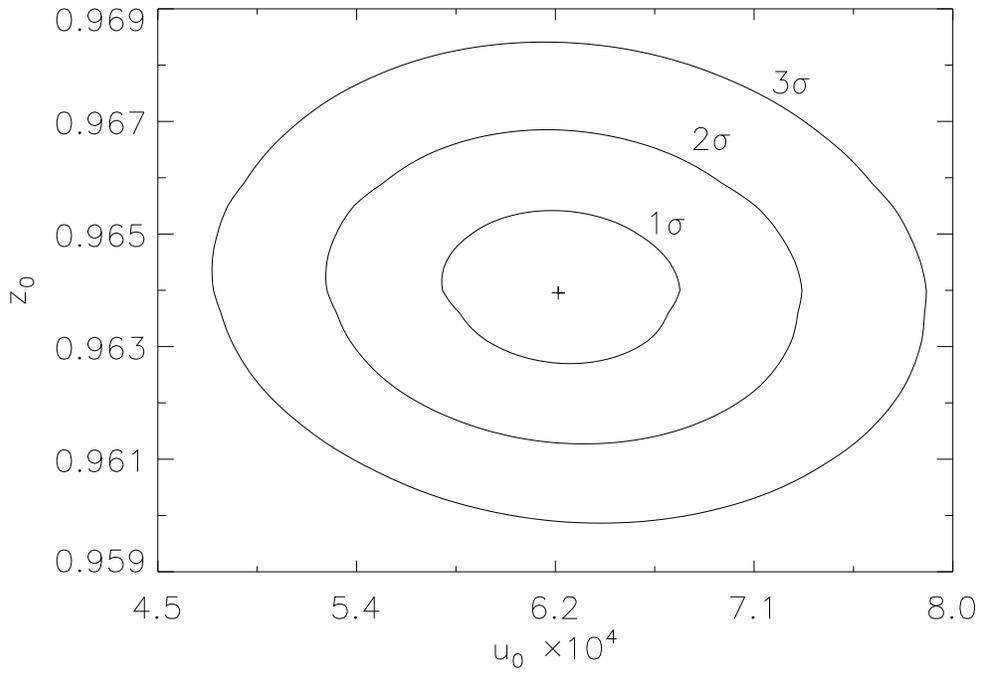}
\caption{$\chi^2$ contours as a function of impact parameter, $u_0$, and $z_0\equiv u_0/\rho_{\star}$ where $\rho_{\star}=\theta_{\star}/\theta_{\rm{E}}$ is the normalized source size. The best fit is marked with a plus sign. \label{fig:u0z0}}
\end{figure}

If the source passes very close to the lens star, finite-source effects will smooth out the peak of the light curve and allow a measurement of the source size $\rho_{\star}$. Although finite-source effects are not obvious from a visual inspection of the light curve, including them yields a dramatic improvement in $\chi^2$. In order to fit for finite source effects, we first estimate the limb-darkening of the source from its color and magnitude. We combine the color and magnitude of the source with the Yale-Yonsei isochrones \citep{Demarque04}, assuming a distance of $D_s=8$ kpc and solar metallicity, to estimate $T_{{\rm eff}}=5250$K and $\log g = 4.5$. We use these values to calculate the limb-darkening coefficients, $u$, from \citet{Claret00}, assuming a microturblent velocity of 2 km/s.  We calculate the linear limb-darkening parameters $\Gamma_V$ and $\Gamma_I$ using $\Gamma=2u/(3-u)$ to find $\Gamma_V=0.65$ and $\Gamma_I=0.47$. We use these values in our finite-source fits to the data. We find that a point-lens fit including finite-source effects is preferred by $\Delta\chi^2$ of 2647.85 over a fit assuming a point source. We search a grid of $u_0$ and $\rho_{\star}$ near the minimum to confirm that this is a well constrained result. We use $z_0 = u_0/\rho_{\star}$ as a proxy for $\rho_{\star}$ following \citet{Yoo04}. The resultant $\chi^2$ map in the $u_0$-$z_0$ plane is shown in Figure \ref{fig:u0z0}. Our best-fit value for $\rho_{\star}$ is $6.6\pm0.6 \times 10^{-4}$. For this value of $\rho_{\star}$, $z_0$ is almost unity, indicating that the source just barely grazed the lens star. The other parameters for our best-fit including finite-source effects are given in Table \ref{tab:params}.

\subsubsection{Source Size}
We convert the dereddened color and magnitude of the source to $(V-K)$ using \citet{Bessell88}, and combine them with the surface brightness relations in \citet{Kervella04} to derive a source size of $\theta_{\star} = 0.54\pm 0.4 \,\mu\mathrm{as}$. The uncertainty in $\theta_{\star}$ comes from two sources: the uncertainty in the flux and the uncertainty in the conversion from the observed $(V-I)$ color to surface brightness. The uncertainty in the flux (i.e. the model fit parameter $f_{s, I}$) is $8.5\%$, and we adopt $7\%$ as the uncertainty due to the surface brightness conversion. From equation (\ref{eqn:rho}), we find that $\theta_{\rm{E}} = \theta_{\star}/\rho_{\star} = 0.81\pm 0.07 \,\mathrm{mas}$. We also calculate the (geocentric) proper motion of the source: $\mu_{\rm geo}=\theta_{\rm E}/t_{\rm E}=2.7 \pm 0.2\,$ mas/yr. Because the peak flux ($\propto f_{s,I}/\rho_{\star}$) and source crossing time ($\rho_{\star}t_{\rm{E}}$) are both essentially direct observables, and so are well constrained by the light curve, the fractional uncertainty in $\theta_{\rm E}$ and $\mu_{\rm geo}$ are comparable to the fractional uncertainty in $\theta_{\star}$. This result is generally applicable to point-lens/finite-source events and is discussed in detail in the Appendix.

\subsection{Parallax}

Given that we have a measurement for $\theta_{\rm{E}}$, if we can also measure microlens parallax, $\pi_{\rm{E}}$, we can combine these quantities to derive the mass of the lens and its distance. The mass of the lens is given by 
\begin{equation}
\label{eqn:mass}
M_l=\frac{\theta_{\rm{E}}}{\kappa\pi_{\rm{E}}}, \quad \kappa \equiv \frac{4G}{c^2 \mathrm{AU}} \simeq 8.14 \frac{\mathrm{mas}}{M_{\odot}} .
\end{equation}
Its distance $D_l$ is
\begin{equation}
\label{eqn:dl}
\frac{1\mathrm{\,AU}}{D_l} = \pi_l = \pi_s+\pi_{\rm{rel}},
\end{equation}
where $\pi_l$ is the parallax of the lens, $\pi_s=0.125\,$ mas is the parallax of the source (assuming a distance of $D_s=8\,$ kpc), and $\pi_{\rm{rel}} = \theta_{\rm{E}}\pi_{\rm{E}}$.

Microlens parallax is the combination of two observable parallax effects in a microlensing event. Terrestrial parallax occurs because observatories located on different parts of the Earth have slightly different lines of sight toward the event and so observe slight differences in the peak magnification and in the timing of the peak, described by the parameters $u_0$ and $t_0$, respectively \citep{Hardy95,Holz96}. Orbital parallax occurs because the Earth moves in its orbit during the event, again, changing the apparent line of sight. \citet{Gould97} argued that one might expect to measure both finite-source effects and terrestrial parallax in extreme high-magnification events. We fit the light curve for both of the sources of parallax, including finite-source effects. Fitting for both kinds of parallax simultaneously yields a $\Delta\chi^2$ improvement of 165 (see Table \ref{tab:params}). We find $\bpi_{\rm E} = (\pi_{\rm{E,E}}, \pi_{\rm{E,N}}) = (-0.15\pm0.02, 0.02\pm0.02)$, where $\pi_{\rm{E,E}}$ and $\pi_{\rm{E,N}}$ are the projections of $\bpi_{\rm E}$ in the East and North directions, respectively.

\citet{Smith03} showed that for orbital parallax and a constant acceleration, $u_0$ has a sign degeneracy. This degeneracy may be broken if terrestrial parallax is observed \citep[see also][]{Gould04}. In the fits described above, we assumed $u_0>0$. We repeat the parallax fit fixing $u_0<0$. We find that the $+u_0$ solution is preferred over the $-u_0$ case by $\Delta\chi^2 = 37$ (see Table \ref{tab:params}). 

We perform a series of fits in order to isolate the source of the parallax signal, i.e. whether it is primarily due to orbital parallax or terrestrial parallax. We first fit the light curve for orbital parallax alone and then fit for terrestrial parallax alone. The results are given in Table \ref{tab:params}. For $+u_0$, the orbital parallax fit gives $(\pi_{\rm{E,E}}, \pi_{\rm{E,N}})=(1.5\pm0.4, -0.3\pm0.2)$ and a $\Delta\chi^2$ improvement of $\sim 16$ over the finite-source fit without parallax. In contrast, the $+u_0$ fit for terrestrial parallax alone yields $\Delta\chi^2=166$ and $(\pi_{\rm{E,E}},\pi_{\rm{E,N}})=(-0.16\pm0.02,0.03\pm0.02)$. While the orbital and terrestrial parallaxes are nominally inconsistent at more than 3$\sigma$, from previous experience \citep{Poindexter05} we know that low-level orbital parallax can be caused by small systematic errors or xallarap (the orbital motion of the source due to a companion), so we ignore this discrepancy. From the $\Delta\chi^2$ values, it is clear that terrestrial parallax dominates the microlens parallax signal in this event, so any spurious orbital parallax signal does not affect our final results.

We also confirm that the terrestrial parallax signal is seen in multiple observatories, and thus cannot be attributed to systematics in a single data set. To test this, we repeat the fits for parallax excluding the data from an individual observatory. If a data set is removed and the parallax becomes consistent with zero, then that observatory contributed significantly to the detection of the signal. Using this process of elimination, we find that the signal comes primarily from the MOA and Bronberg data sets.

Given the results of these various fits, we conclude that the best fit to the data is for the $+u_0$ solution, and we include both forms of parallax for internal consistency. Combining this parallax measurement with our measurement of $\theta_{\rm{E}}$ from $\S$\ref{sec:source}, we find $M_l = 0.64 \pm 0.1 M_{\odot}$ and $D_l=4.0 \pm 0.6 \,\mathrm{kpc}$ ($\pi_{\rm{rel}} = 0.13 \pm 0.02 \,\mathrm{mas}$) using equations (\ref{eqn:mass}) and (\ref{eqn:dl}). 

\section{The Blended Light}
\label{sec:blend}

\begin{figure}
\includegraphics[width=5.5in]{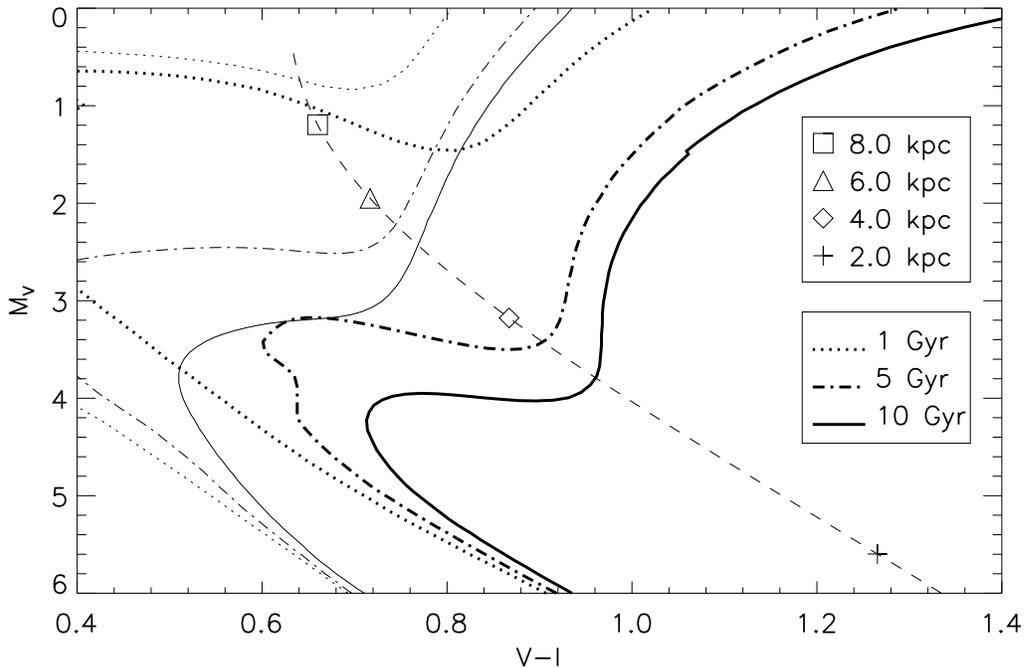}
\caption{Possible absolute magnitudes and colors for the blend plotted with Yale-Yonsei isochrones \citep{Demarque04}.  The isochrones plotted are the Y$^2$ isochrones for solar (thick) and sub-solar metallicities (thin) for populations 1 (dotted), 5 (dot-dashed), and 10 Gyr old (solid). The dashed line shows the color and magnitude of the blend for a continuous distribution of distances assuming a dust model that decreases exponentially with scale height. The square shows the absolute magnitude and color of the blend assuming it has the same distance (8 kpc) and reddening as the clump. The plus sign, diamond, and triangle show the absolute magnitude and color using our simple dust model and distances of 2, 4, and 6 kpc, respectively. If the blend is a companion to the lens, it would be at a distance of 4 kpc (diamond). \label{fig:yyiso}}
\end{figure}

The centroid of the light at baseline when the source is faint is different from the centroid at peak magnification, indicating that light from a third star is blended into the PSF. The measured color and magnitude of blended light are $[I, (V-I)]_b = [17.21\pm 0.01, 2.32\pm 0.02]$. Stars of this magnitude are relatively rare, and so the most plausible initial guess is that the third star is either a companion to the source or a companion to the lens. If the former, we can use the values of $A_I$ and $E(V-I)$ we found above to derive the intrinsic color of the blend: $[I, (V-I)]_{0,b} = [15.05, 0.66]$. This assumes that the blend is in the bulge at a distance of 8 kpc, giving an absolute magnitude of $M_{I,b} = 0.53$ and $M_{V,b}=1.19$. Figure \ref{fig:yyiso} shows this point (open square) compared to solar (Z=0.02) and sub-solar metallicity (Z=0.001) Yale-Yonsei isochrones at 1, 5, and 10 Gyrs \citep{Demarque04}. These isochrones show that the values of $[M_V, (V-I)_0]_{b}$ may be consistent with a sub-giant that is a couple Gyr old, but a more precise determination of age is not possible since the age is degenerate with the unknown metallicity of the blend.

If the blend is a companion to the lens, however, it lies in front of some fraction of the dust. In order to derive a dereddened color and absolute magnitude to this star, we need a model for the dust. We explore this scenario using a simple model for the extinction that is constant in the plane of the disk and decreases exponentially out of the plane with a scale height of $H_0 = 100\,$ pc:
\begin{equation}
\label{eqn:aiofd}
A_I(d) = K_1\left[1-\exp\left(\frac{-D\sin b}{H_0}\right)\right],
\end{equation}
where $D$ is the distance to a given point along the line of sight, $b$ is the Galactic latitude, and $K_1$ is a constant. We can solve for $K_1$ by substituting in the value of $A_I$ that we find for the source at 8 kpc. We then model the selective extinction in a similar manner:
\begin{equation}
\label{eqn:eviofd}
E(V-I) = K_2\left[1-\exp\left(\frac{-D\sin b}{H_0}\right)\right],
\end{equation}
and solve for $K_2$ using the value of E(V-I) calculated for the source at 8 kpc. From equations \ref{eqn:aiofd} and \ref{eqn:eviofd}, we can recover the intrinsic color and magnitude of the blend assuming it is at various distances. In Figure \ref{fig:yyiso}, we plot a point assuming the blend is at the distance of the lens, 4.0 kpc. By interpolating the isochrones and assuming a solar metallicity, we find that the blend is consistent with being a 1.4 $M_{\odot}$ sub-giant companion to the lens with an age of 3.8 Gyr. For comparison, we also plot a line showing how the inferred color and magnitude of the blend vary with the assumed distance.

\subsection{Astrometric Offset}
From the measured blend flux, one can determine the astrometric offset of the source and blend by comparing the centroid of light during and after the event. At a given epoch, the centroid is determined by the ratio of the flux of the blend to the sum of the fluxes of the source and lens. That ratio depends on the magnification of the source. Thus, if we know the magnification of the source at two different epochs and the intrinsic magnitude of the source and the blend, we can solve for the separation of the lens and the blend. We find $\Delta\theta = 153 \pm 18 \,\mathrm{mas}$. Given this offset, we will show below that based on the lack of shear observed in the light curve, the blended light cannot lie far in the foreground and thus cannot be the sub-giant companion to the lens hypothesized above.

\subsection{Search for Shear}
\label{sec:shear}

\begin{figure}
\includegraphics[width=5.5in]{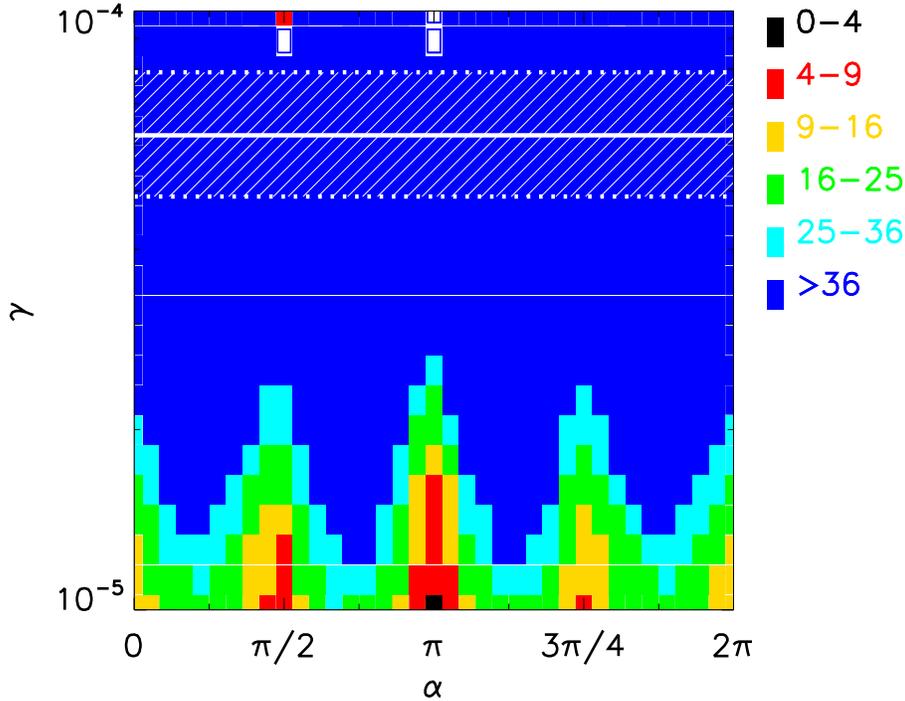}
\caption{Shear as a function of $\alpha$ (angular position with respect to the motion of the source). Open symbols indicate an improved $\chi^2$ compared to the finite-source point-lens fit. Filled symbols indicate a worse fit. The magnitude of $\Delta\chi^2$ is indicated by the color legend shown. The solid line indicates our calculated value for the shear assuming the blend is at the same distance as the lens. The shaded area shows the $1\,\sigma$ limits on this value from the uncertainty in the centroid of the PSF (see text). \label{fig:shear}}
\end{figure}

Because all stars have gravity, if the blend described above lies between the observer and the source, it will induce a shear $\gamma$ in the light curve. We can estimate the size of the shear using the observed astrometric offset and assuming that the blend is a $1.4 M_{\odot}$ companion to the lens.
\begin{eqnarray}
\gamma &=& \frac{\theta_{{\rm E},b}^2}{\Delta\theta^2} = \frac{\kappa\pi_{{\rm rel},b} M_{b}}{\Delta\theta^2}, \nonumber\\
&=&6.2\times 10^{-5} \left(\frac{\pi_{{\rm rel}}}{0.13\, \mathrm{mas}}\right)\left(\frac{M_{b}}{1.4\, M_{\odot}}\right)\left(\frac{\Delta\theta}{153\, \mathrm{mas}}\right)^{-2}.\label{eqn:shear}
\end{eqnarray}
Using the $1\,\sigma$ upper limit on the separation (171 mas), we find a minimum shear of $\gamma=4.9\times10^{-5}$ if the blend is a companion to the lens. To determine if this value is consistent with the light curve, we perform a series of fits to the data using binary-lens models that cover a wide range of potential shears. The effect of the shear is to introduce two small bumps into the light curve as the small binary caustic crosses the limb of the source, and this is indeed what we see in the binary-lens models we calculate.

Because the separation between the lens and a companion is large ($B=\Delta\theta/\theta_{\rm E}\gg 1$), the shear can be approximated as $\gamma \simeq Q/B^2$, where $Q=M_{b}/M_l$ is the mass ratio of the companion and the lens. This reduces the number of parameters that need to be considered from three to two: $\gamma$ and $\alpha$, the angular position of the blend with respect to the motion of the source. We use a grid search of $\gamma$ and $\alpha$ to place limits on the shear. For each combination of $\gamma$ and $\alpha$, we generate a binary light curve in the limit $B\gg1$ that satisfies $Q=\gamma B^2$ and fit it to the data using a Markov Chain Monte Carlo with 1000 links. We bin the data over the peak to reduce computing time. We compute the difference in $\chi^2$ between the binary model and the best-fit finite-source point-lens model. Figure \ref{fig:shear} shows the results of the grid search over-plotted with the upper and lower limits on the shear assuming the blend is a companion to the lens. From this figure, we infer that a shear of $6.2\times10^{-5}$ is inconsistent with our data since it is in a region where the fit is worse by $\Delta\chi^2>36$. 

The two minima in the $\chi^2$ map at $\gamma \sim10^{-4}, \alpha=\pi/2, \pi$ are well-defined but appear to be due to a single, deviant data point. Fits to the data with these binary models show improvement in the fit to this data point, but the residuals from these fits for the other data points are large and show increased structure. Thus, we believe these minima to be spurious and conclude that the maximum shear that is consistent with our data ($\Delta\chi^2 \le 9$) is $\gamma_{\mathrm{max}} = 1.6 \times 10^{-5}$.

Since we have ruled out the scenario where the blend is a companion to the lens, we need to ask what possible explanations for the blend are consistent both with $\gamma_{\mathrm{max}}$ and with the observed color and magnitude. Given $\gamma_{\mathrm{max}}$, we can place constraints on the distance to the blend, $D_{b}$, for a given mass. The distance is given by
\begin{eqnarray}
& & D_{b} = \frac{1}{\pi_{b}},\\
& & \mathrm{where}\quad \pi_{b} = \pi_s +\pi_{{\rm rel},b} = \pi_s + \frac{\gamma(\Delta\theta)^2}{\kappa M_{b}}
\end{eqnarray}
If we assume $M_b=1 M_{\odot}$, $\gamma=\gamma_{\mathrm{max}}$, and use previously stated values for the other parameters, we find $D_b > 5.8\,$ kpc. A metal-poor sub-giant with this mass located at or beyond this distance would be consistent with the observed color and magnitude of the blend given the simple extinction model described above. However, other explanations are also possible. For example, if the mass of the blend were decreased, $\pi_{b}$ would increase, and a slightly closer distance would be permitted. Thus, we cannot definitively identify the source of the blended light. However, given that $\gamma_{\mathrm{max}}$ is very small, we can ignore any potential shear contribution in later analysis.

\section{Limits on Planets}
\label{sec:planets}

\begin{figure}
\includegraphics[width=5.5in]{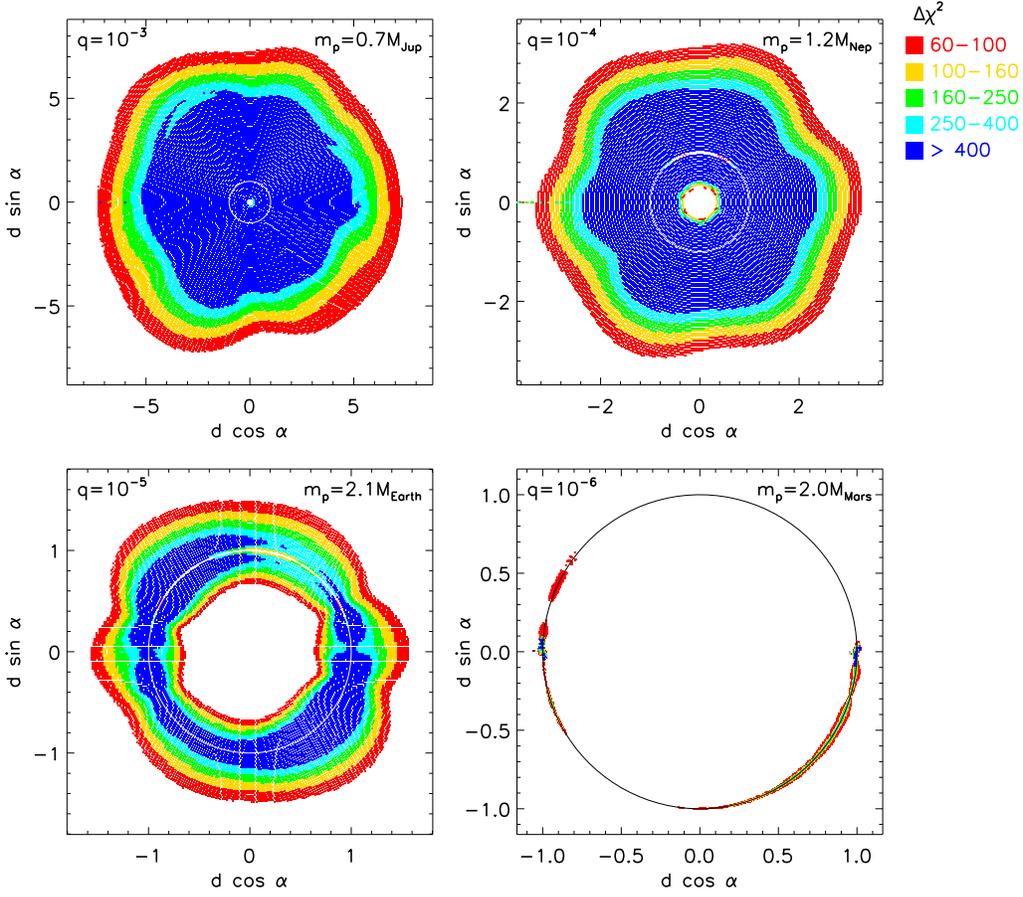}
\caption{Planet sensitivity as a function of distance from the lens in units of Einstein radii. The white/black circle indicates the Einstein ring ($d=1$). The mass ratios and corresponding planet masses are indicated on each plot. The colors indicate the $\Delta\chi^2$ that would be caused by a planet at that location. \label{fig:maps}}
\end{figure}

\begin{figure}
\includegraphics[width=5.5in]{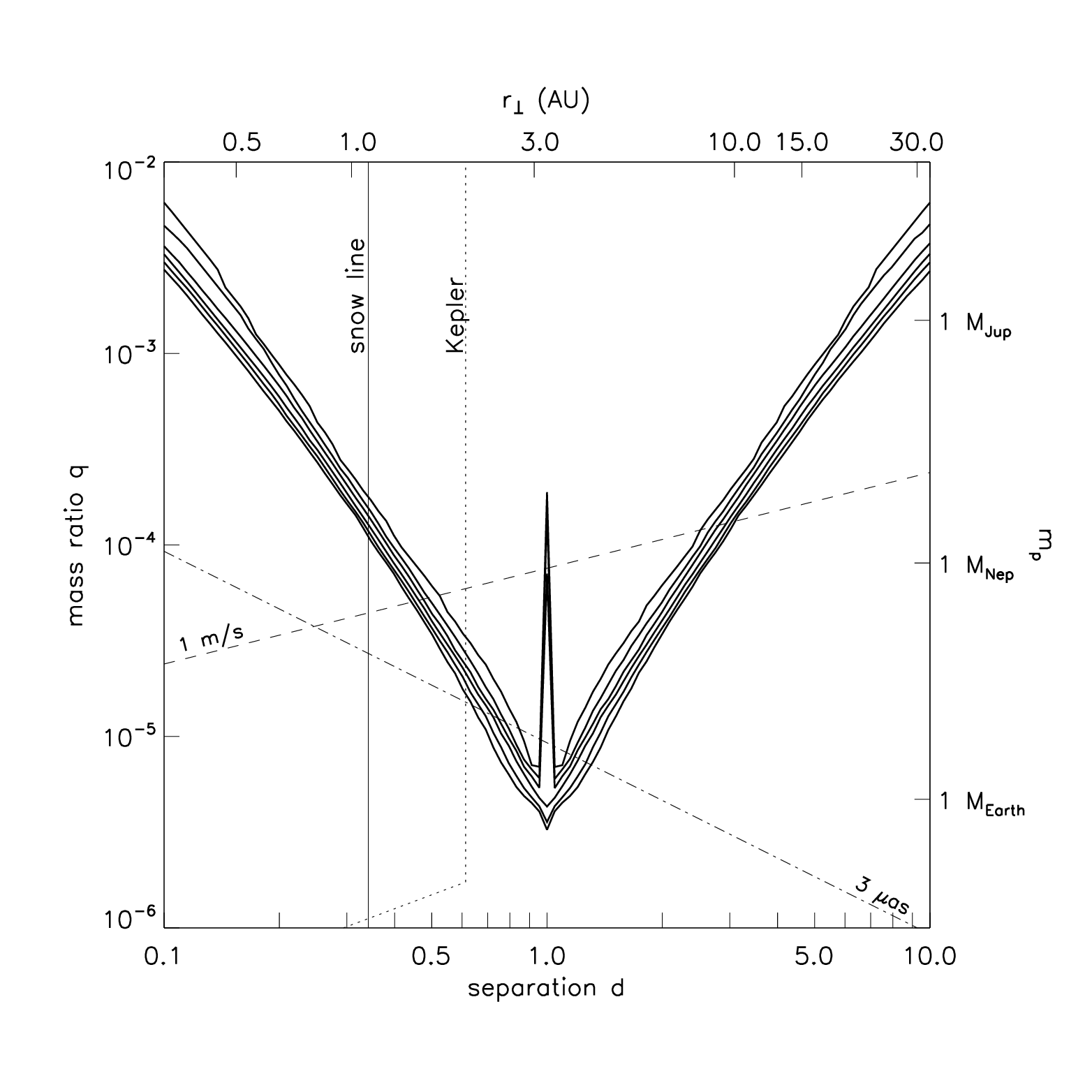}
\caption{Detection efficiency map in the $(d,q)$ plane, i.e. projected separation in units of $\theta_{\rm E}$ and planet-star mass ratio. The contours show detection efficiencies of 0.99, 0.90, 0.75, 0.50, 0.25, and 0.10 from inside to outside. The inner spike is due to resonant caustic effects at the Einstein ring. The upper and right axes translate $(d,q)$ into physical units ($r_{\perp}, m_p$), i.e. physical projected separation and planet mass. The vertical solid line shows the position of the snow line for this star. The dotted line shows Kepler's sensitivity to planets around the lens star assuming $m_V=12$. The cutoff in separation ($d \simeq 0.6$) occurs where a planet's orbital period is equal to Kepler's mission lifetime of 3.5 yrs. The dashed line shows the sensitivity limit for radial velocity observations with 1 m/s precision. The dot-dashed line shows the sensitivity limit for a space-based astrometry mission with precision of 3\,$\mu$as assuming the star is at 10 pc.\label{fig:bqeff}}
\end{figure}

We use the method described by \citet{Rhie00} to quantify the sensitivity of this event to planets. This approach is used for events such as this one for which the residuals are consistent with a point-lens. Rather than fitting binary models for planetary companions to our data as advocated by \citet{Gaudi00}, we generate a binary model from the data and fit it with a point-lens model. When the single-lens parameters are well constrained (as is the case with
OGLE-2008-BLG-279), these two approaches are essentially equivalent (see
the discussion in \citealt{Gaudi02} and \citealt{Dong06}). We create a magnification map assuming an impact parameter, $d$, and star/planet mass ratio, $q$, using a lens with the characteristics from our finite-source fit. The method for creating the magnification map is described in detail in \citet{Dong06} and \citet{Dong09}. For each epoch of our data, we generate a magnification due to the binary lens assuming some position angle, $\alpha$, of the source's trajectory relative to the axis of the binary and assign it the uncertainty of the datum at that epoch. As in $\S$\ref{sec:shear}, we use binned data for this analysis.

For $q = 10^{-3}, 10^{-4}, 10^{-5}, \,\mathrm{and }\, 10^{-6}$ we search a grid of $d$, $\alpha$ and compute the $\Delta\chi^2$. Based on the systematics in our data, we choose a threshold $\Delta\chi^2_{\mathrm{min}}=160$ \citep{Gaudi00}. For $\Delta\chi^2 > \Delta\chi^2_{\mathrm{min}}$, the fit is excluded by our data, and we are sensitive to a planet of mass ratio $q$ at that location. We repeat the analysis using unbinned data for a small subset of points and confirm that the $\Delta\chi^2$ for fits with the unbinned data is comparable to fits with binned data. Figure \ref{fig:maps} shows the sensitivity maps for four values of $q$. These maps show good sensitivity to planets with mass ratios $q = 10^{-3}, 10^{-4}, \,{\rm and}\, 10^{-5}$ and some sensitivity to planets with $q=10^{-6}$. For our measured value of $M_{l} =  0.64 M_{\odot}$, a mass ratio of $q=10^{-3}$ corresponds to a planet mass $m_p = 0.67 M_{\mathrm{Jup}}$ and a mass ratio of $q=10^{-6}$ corresponds to $m_p \simeq 2 M_{\mathrm{Mars}}$. The results bear a striking resemblance to the hypothetical planet sensitivity of the $A_{{\rm max}}\sim3000$ event OGLE-2004-BLG-343 if it had been observed over the peak \citep{Dong06}. In particular, this event shows nearly uniform sensitivity to planets at all angles $\alpha$ for large mass ratios. The hexagonal shape of the sensitivity map is the imprint of the difference between the magnification maps of planetary-lens models and their corresponding single-lens models (see upper panel of Fig. 3 in \citet{Dong09}).

Figure \ref{fig:bqeff} shows a map of the planet detection efficiency for this event. The efficiency is the percentage of trajectories, $\alpha$, at a given mass ratio and separation that have $\Delta\chi^2>\Delta\chi^2_{\mathrm{min}}$ \citep{Gaudi00}. The efficiency contours are all quite close together because of the angular symmetry described above for the planet sensitivity maps. Because we measure the distance to the lens, we know the projected separation, $r_{\perp}$, in physical units:
\begin{equation}
r_{\perp}=d\theta_{\rm{E}} D_l .
\label{eqn:b}
\end{equation}
Since we know $M_l$, we also know the planet mass, $m_p = q M_l$. We can rule out Neptune-mass planets with projected separations of 1.5--7.2 AU ($d=0.5$--2.2) and Jupiter-mass planets with separations of 0.54--19.5 AU ($d=0.2$--6.0). We are also able to detect Earth-mass planets near the Einstein ring, although the efficiency is low. The region where this event is sensitive to giant planets probes well beyond the snow line of this star, which we estimate to be at 1.1 AU assuming $a_{{\rm snow}}=2.7 {\rm AU} (M_{\star}/M_{\odot})^{2}$ \citep{Ida04}. The observed absence of planets, especially Neptunes, immediately beyond the snow line of this star is interesting given that core-accretion theory predicts that Neptune-mass planets should preferentially form around low-mass stars \citep{Laughlin04, Ida05}. 

It is also interesting to consider how the sensitivity of this event to planets compares to the sensitivity of other planet-search techniques. Obviously, because of the long timescales involved, most transit searches barely probe the region of sensitivity for this event. As a space-based mission, the Kepler satellite has the best opportunity to probe some of the microlensing parameter space using transits. Using equation 21 from \citet{Gaudi07}, we can estimate Kepler's sensitivity to transits around this star:
\begin{equation}
m_p = 0.22\left(\frac{S/N}{10}\right)^{3/2}\left(\frac{a}{1\,\mathrm{AU}}\right)^{3/4}10^{0.3(m_V-12)} M_{\mathrm{Earth}},
\end{equation}
where $(S/N)$ is the signal-to-noise ratio, $a$ is the semi-major axis of the planet, and $m_V$ is the apparent magnitude of the star. We have assumed that the density of the planet is the same as the density of the Earth and the stellar mass-radius relation $R_{\star}=kM_{\star}^{0.8}$ \citep{Cox00}. Kepler is also limited by its mission lifetime of 3.5 yrs. For periods longer than this, it becomes increasingly unlikely that Kepler will observe a transit \citep{Yee08}. This limits the sensitivity to planets within $\sim2$\,AU where the period is less than the mission lifetime. These boundaries are plotted in Figure \ref{fig:bqeff}.

For comparison, we can also estimate the sensitivity of the radial velocity technique to planets around a star of this mass assuming circular orbits and an edge-on system. Radial velocity is sensitive to planets of mass
\begin{equation}
m_p = 8.9 \left(\frac{\sigma_{\rm{RV}}}{1\,\mathrm{m/s}}\right)\left(\frac{S/N}{10}\right)\left(\frac{N}{100}\right)^{-1/2}\left(\frac{a}{1\,\mathrm{AU}}\right)^{1/2} M_{\mathrm{Earth}},
\end{equation}
where $\sigma_{\rm{RV}}$ is the precision, and $N$ is the number of observations.  The limit of radial velocity sensitivity is plotted in Figure \ref{fig:bqeff} as a function of separation assuming a precision of 1 m/s. Additionally, we can consider how this microlensing event compares to the sensitivity of a space-based astrometry mission with microarcsecond precision ($\sigma_a = 3\, \mu\mathrm{as}$):
\begin{equation}
m_p = 6.4 \left(\frac{\sigma_a}{3\,\mu\mathrm{as}}\right)\left(\frac{S/N}{10}\right)\left(\frac{N}{100}\right)^{-1/2}\left(\frac{a}{1\,\mathrm{AU}}\right)^{-1}\left(\frac{d}{10\,\mathrm{pc}}\right) M_{\mathrm{Earth}}.
\end{equation}
We assume circular face-on orbits. We show the limiting mass as a function of semi-major axis in Figure \ref{fig:bqeff} for 3 $\mu$as precision. While these contours encompass a large region of the parameter space, they do not take into account the time it takes to make the observations, which increases with increasing semi-major axis. Furthermore, we only expect this kind of astrometric precision from a future space mission, whereas this event shows that microlensing is currently capable of finding these planets from the ground. This discussion shows that microlensing is sensitive to planets in regions not probed by transits and radial velocity and will be particularly important for finding planets at wide separations where the periods are long. For example, for semi-major axis $a=4\,$AU (near the maximum sensitivity shown in Fig. \ref{fig:bqeff}), the period is $P\simeq 10\,$yr.

\section{Summary}
\label{sec:conclusions}
The extreme magnification microlensing event OGLE-2008-BLG-279 allowed us to place broad constraints on planets around the lens star. Even with a more conservative detection threshold ($\Delta\chi^2>160$), this event is more sensitive than any previously analyzed event (the prior record holder was MOA-2003-BLG-32; \citealt{Abe04}). Furthermore, because we observe both parallax and finite-source effects in this event, we are able to measure the mass and distance of an isolated star ($M_l = 0.64 \pm 0.10 M_{\odot}, D_l = 4.0 \pm 0.6\rm{kpc}$). Using these properties of the lens star, we convert the mass ratio and projected separation to physical units. We can exclude giant planets around the lens star in the entire region where they are expected to form, out beyond the snow line. For example, Jupiter-mass planets are excluded from 0.54--19.5 AU. Events like this that can detect or exclude a broad range of planetary systems out beyond the snow line will be important for determining the planet frequency at large separations and constraining models of planet formation and migration.

\acknowledgements{We acknowledge the following support: NSF AST-0757888 (AG,SD,JCY); NASA NNG04GL51G (DD,AG,RP); Polish MNiSW N20303032/4275 (AU); Korea Astronomy and Space Science Institute (B-GP,C-UL);Creative Research Initiative Program (2009-008561) of  Korea Science and Engineering Foundation (CH).}

\section{Appendix: Uncertainty in $\theta_{\star}$ $\mu$, and $\theta_{\rm E}$}

In the present case, the fractional errors in $\theta_{\star}$, $\mu$, and
$\theta_{\rm E}$ are all very nearly the same, although for somewhat
different reasons.  Since the same convergence of errors is likely
to occur in many point-lens/finite-source events, we briefly
summarize why this is the case.  We first write (generally),
$$
\theta_{\star} = \sqrt{f_s}/Z
$$
where $f_s$ is the source flux as 
determined from the model, and $Z$ is the remaining set of factors, which generally include the surface brightness of the source, uncertainties due to the calibration of the source flux, and numerical constants.  
Next, we write
$$
\mu = {\theta_{\rm E}\over t_{\rm E}} = {\theta_{\star}\over t_{\star}} = {\sqrt{f_s}\over Z}\,{1\over t_{\star}}
\qquad
\theta_{\rm E} = {\theta_{\star}\over\rho} = {1\over Z\sqrt{f_s}}\,f_{\rm grand}
$$
where $f_{\rm grand}\equiv f_s/\rho$ and $t_{\star}\equiv \rho t_{\rm E}$.  
We note that for point-lens events with strongly detected finite
source effects, $t_{\star}$ and $f_{\rm grand}$ are quasi-observables,
and so have extremely small errors. For example, if $u_0=0$,
then $2t_{\star}$ is just the observed source crossing time while
$2f_{\rm grand}[1+(3\pi/8-1)\Gamma]$ is the observed peak flux.
Even for $u_0\not=0$, these quantities are very strongly constrained,
with errors $\sigma_{f_{\rm grand}} = 0.4\%$ and $\sigma_{t_{\star}}=0.3\%$ in the present case.  Since the errors
in $f_s$ and $Z$ are independent, the fractional errors in
$\theta_{\star}$, $\mu$, and $\theta_{\rm E}$ are each equal to $[(1/4)(\sigma_{f_s}/f_s)^2 + (\sigma_Z/Z)^2]^{1/2}$.  In the present case, $\sigma_{f_s}/f_s$ is
given by the fitting code to be 8.5\%, while we estimate $\sigma_Z/Z$
to be 7\%, and therefore find a net error in all three quantities ($\theta_*,\theta_{\rm E}$, and $\mu$) of
8\%.

\bibliographystyle{apj}

\begin{thebibliography}
\expandafter\ifx\csname natexlab\endcsname\relax\def\natexlab#1{#1}\fi
\bibitem[{{Abe} {et~al.}(2004){Abe}, {Bennett}, {Bond}, {Eguchi}, {Furuta},
  {Hearnshaw}, {Kamiya}, {Kilmartin}, {Kurata}, {Masuda}, {Matsubara},
  {Muraki}, {Noda}, {Okajima}, {Rakich}, {Rattenbury}, {Sako}, {Sekiguchi},
  {Sullivan}, {Sumi}, {Tristram}, {Yanagisawa}, {Yock}, {Gal-Yam}, {Lipkin},
  {Maoz}, {Ofek}, {Udalski}, {Szewczyk}, {{\.Z}ebru{\'n}}, {Soszy{\'n}ski},
  {Szyma{\'n}ski}, {Kubiak}, {Pietrzy{\'n}ski}, \& {Wyrzykowski}}]{Abe04}
{Abe}, F., et al. 2004, Science, 305, 1264

\bibitem[{{Batista} {et~al.}(2009){Batista}, {Dong}, {Gould}, {Beaulieu},
  {Cassan}, {Christie}, {Han}, {Udalski}, {Allen}, {DePoy}, {Gal-Yam}, {Gaudi},
  {Johnson}, {Kaspi}, {Lee}, {Maoz}, {McCormick}, {McGreer}, {Monard},
  {Natusch}, {Ofek}, {Park}, {Pogge}, {Polishook}, {Shporer}, {Albrow},
  {Bennett}, {Brillant}, {Bode}, {Bramich}, {Burgdorf}, {Caldwell}, {Calitz},
  {Cole}, {Cook}, {Coutures}, {Dieters}, {Dominik}, {Dominis Prester},
  {Donatowicz}, {Fouqu{\'e}}, {Greenhill}, {Hoffman}, {Horne}, {J{\o}rgensen},
  {Kains}, {Kane}, {Kubas}, {Marquette}, {Martin}, {Meintjes}, {Menzies},
  {Pollard}, {Sahu}, {Snodgrass}, {Steele}, {Tsapras}, {Wambsganss},
  {Williams}, {Zub}, {Wyrzykowski}, {Kubiak}, {Szyma{\'n}ski},
  {Pietrzy{\'n}ski}, {Soszy{\'n}ski}, {Szewczyk}, {Ulaczyk}, {Abe}, {Bond},
  {Fukui}, {Furusawa}, {Hearnshaw}, {Holderness}, {Itow}, {Kamiya},
  {Kilmartin}, {Korpela}, {Lin}, {Ling}, {Masuda}, {Matsubara}, {Miyake},
  {Muraki}, {Nagaya}, {Ohnishi}, {Okumura}, {Perrott}, {Rattenbury}, {Saito},
  {Sako}, {Skuljan}, {Sullivan}, {Sumi}, {Sweatman}, {Tristram}, \&
  {Yock}}]{Batista09}
{Batista}, V., et al. 2009, ArXiv e-prints

\bibitem[{{Beaulieu} {et~al.}(2006){Beaulieu}, {Bennett}, {Fouqu{\'e}},
  {Williams}, {Dominik}, {J{\o}rgensen}, {Kubas}, {Cassan}, {Coutures},
  {Greenhill}, {Hill}, {Menzies}, {Sackett}, {Albrow}, {Brillant}, {Caldwell},
  {Calitz}, {Cook}, {Corrales}, {Desort}, {Dieters}, {Dominis}, {Donatowicz},
  {Hoffman}, {Kane}, {Marquette}, {Martin}, {Meintjes}, {Pollard}, {Sahu},
  {Vinter}, {Wambsganss}, {Woller}, {Horne}, {Steele}, {Bramich}, {Burgdorf},
  {Snodgrass}, {Bode}, {Udalski}, {Szyma{\'n}ski}, {Kubiak}, {Wi{\c e}ckowski},
  {Pietrzy{\'n}ski}, {Soszy{\'n}ski}, {Szewczyk}, {Wyrzykowski},
  {Paczy{\'n}ski}, {Abe}, {Bond}, {Britton}, {Gilmore}, {Hearnshaw}, {Itow},
  {Kamiya}, {Kilmartin}, {Korpela}, {Masuda}, {Matsubara}, {Motomura},
  {Muraki}, {Nakamura}, {Okada}, {Ohnishi}, {Rattenbury}, {Sako}, {Sato},
  {Sasaki}, {Sekiguchi}, {Sullivan}, {Tristram}, {Yock}, \&
  {Yoshioka}}]{Beaulieu06}
{Beaulieu}, J.-P., et al. 2006, \nat, 439, 437

\bibitem[{{Bennett} {et~al.}(2008){Bennett}, {Bond}, {Udalski}, {Sumi}, {Abe},
  {Fukui}, {Furusawa}, {Hearnshaw}, {Holderness}, {Itow}, {Kamiya}, {Korpela},
  {Kilmartin}, {Lin}, {Ling}, {Masuda}, {Matsubara}, {Miyake}, {Muraki},
  {Nagaya}, {Okumura}, {Ohnishi}, {Perrott}, {Rattenbury}, {Sako}, {Saito},
  {Sato}, {Skuljan}, {Sullivan}, {Sweatman}, {Tristram}, {Yock}, {Kubiak},
  {Szyma{\'n}ski}, {Pietrzy{\'n}ski}, {Soszy{\'n}ski}, {Szewczyk},
  {Wyrzykowski}, {Ulaczyk}, {Batista}, {Beaulieu}, {Brillant}, {Cassan},
  {Fouqu{\'e}}, {Kervella}, {Kubas}, \& {Marquette}}]{Bennett08}
{Bennett}, D.~P., et al. 2008, \apj, 684, 663

\bibitem[{{Bessell} \& {Brett}(1988)}]{Bessell88}
{Bessell}, M.~S. \& {Brett}, J.~M. 1988, \pasp, 100, 1134
\bibitem[{{Bond} {et~al.}(2002){Bond}, {Rattenbury}, {Skuljan}, {Abe}, {Dodd},
  {Hearnshaw}, {Honda}, {Jugaku}, {Kilmartin}, {Marles}, {Masuda}, {Matsubara},
  {Muraki}, {Nakamura}, {Nankivell}, {Noda}, {Noguchi}, {Ohnishi}, {Reid},
  {Saito}, {Sato}, {Sekiguchi}, {Sullivan}, {Sumi}, {Takeuti}, {Watase},
  {Wilkinson}, {Yamada}, {Yanagisawa}, \& {Yock}}]{Bond02}
{Bond}, I.~A., et al. 2002, \mnras, 333, 71

\bibitem[{{Bond} {et~al.}(2004){Bond}, {Udalski}, {Jaroszy{\'n}ski},
  {Rattenbury}, {Paczy{\'n}ski}, {Soszy{\'n}ski}, {Wyrzykowski},
  {Szyma{\'n}ski}, {Kubiak}, {Szewczyk}, {{\.Z}ebru{\'n}}, {Pietrzy{\'n}ski},
  {Abe}, {Bennett}, {Eguchi}, {Furuta}, {Hearnshaw}, {Kamiya}, {Kilmartin},
  {Kurata}, {Masuda}, {Matsubara}, {Muraki}, {Noda}, {Okajima}, {Sako},
  {Sekiguchi}, {Sullivan}, {Sumi}, {Tristram}, {Yanagisawa}, \&
  {Yock}}]{Bond04}
{Bond}, I.~A., et al. 2004, \apjl, 606, L155

\bibitem[{{Claret}(2000)}]{Claret00}
{Claret}, A. 2000, \aap, 363, 1081

\bibitem[{{Cox} (2000)}]{Cox00}
  {Cox}, A.~N. 2000, in Allen's Astrophysical Quantities (New York: AIP), 389

\bibitem[{{Demarque} {et~al.}(2004){Demarque}, {Woo}, {Kim}, \&
  {Yi}}]{Demarque04}
{Demarque}, P., {Woo}, J.-H., {Kim}, Y.-C., \& {Yi}, S.~K. 2004, \apjs, 155,
  667

\bibitem[{{Dong} {et~al.}(2006){Dong}, {DePoy}, {Gaudi}, {Gould}, {Han},
  {Park}, {Pogge}, {Udalski}, {Szewczyk}, {Kubiak}, {Szyma{\'n}ski},
  {Pietrzy{\'n}ski}, {Soszy{\'n}ski}, {Wyrzykowski}, \&
  {{\.Z}ebru{\'n}}}]{Dong06}
{Dong}, S., et al. 2006, \apj, 642, 842

\bibitem[{{Dong} {et~al.}(2009){Dong}, {Bond}, {Gould}, {Koz{\l}owski},
  {Miyake}, {Gaudi}, {Bennett}, {Abe}, {Gilmore}, {Fukui}, {Furusawa},
  {Hearnshaw}, {Itow}, {Kamiya}, {Kilmartin}, {Korpela}, {Lin}, {Ling},
  {Masuda}, {Matsubara}, {Muraki}, {Nagaya}, {Ohnishi}, {Okumura}, {Perrott},
  {Rattenbury}, {Saito}, {Sako}, {Sato}, {Skuljan}, {Sullivan}, {Sumi},
  {Sweatman}, {Tristram}, {Yock}, {The MOA Collaboration}, {Bolt}, {Christie},
  {DePoy}, {Han}, {Janczak}, {Lee}, {Mallia}, {McCormick}, {Monard}, {Maury},
  {Natusch}, {Park}, {Pogge}, {Santallo}, {Stanek}, {The {$\mu$}FUN
  Collaboration}, {Udalski}, {Kubiak}, {Szyma{\'n}ski}, {Pietrzy{\'n}ski},
  {Soszy{\'n}ski}, {Szewczyk}, {Wyrzykowski}, {Ulaczyk}, \& {The OGLE
  Collaboration}}]{Dong09}
{Dong}, S., et al. 2009, \apj, 698, 1826

\bibitem[{{Gaudi} \& {Sackett}(2000)}]{Gaudi00}
{Gaudi}, B.~S. \& {Sackett}, P.~D. 2000, \apj, 528, 56

\bibitem[{{Gaudi} \& {Winn}(2007)}]{Gaudi07}
{Gaudi}, B.~S. \& {Winn}, J.~N. 2007, \apj, 655, 550

\bibitem[{{Gaudi} {et~al.}(2002){Gaudi}, {Albrow}, {An}, {Beaulieu},
  {Caldwell}, {DePoy}, {Dominik}, {Gould}, {Greenhill}, {Hill}, {Kane},
  {Martin}, {Menzies}, {Naber}, {Pel}, {Pogge}, {Pollard}, {Sackett}, {Sahu},
  {Vermaak}, {Vreeswijk}, {Watson}, \& {Williams}}]{Gaudi02}
{Gaudi}, B.~S., et al. 2002,
  \apj, 566, 463

\bibitem[{{Gaudi} {et~al.}(2008){Gaudi}, {Bennett}, {Udalski}, {Gould},
  {Christie}, {Maoz}, {Dong}, {McCormick}, {Szyma{\'n}ski}, {Tristram},
  {Nikolaev}, {Paczy{\'n}ski}, {Kubiak}, {Pietrzy{\'n}ski}, {Soszy{\'n}ski},
  {Szewczyk}, {Ulaczyk}, {Wyrzykowski}, {DePoy}, {Han}, {Kaspi}, {Lee},
  {Mallia}, {Natusch}, {Pogge}, {Park}, {Abe}, {Bond}, {Botzler}, {Fukui},
  {Hearnshaw}, {Itow}, {Kamiya}, {Korpela}, {Kilmartin}, {Lin}, {Masuda},
  {Matsubara}, {Motomura}, {Muraki}, {Nakamura}, {Okumura}, {Ohnishi},
  {Rattenbury}, {Sako}, {Saito}, {Sato}, {Skuljan}, {Sullivan}, {Sumi},
  {Sweatman}, {Yock}, {Albrow}, {Allan}, {Beaulieu}, {Burgdorf}, {Cook},
  {Coutures}, {Dominik}, {Dieters}, {Fouqu{\'e}}, {Greenhill}, {Horne},
  {Steele}, {Tsapras}, {Chaboyer}, {Crocker}, {Frank}, \&
  {Macintosh}}]{Gaudi08}
{Gaudi}, B.~S., et al. 2008, Science, 319, 927

\bibitem[{{Gould}(1997)}]{Gould97}
{Gould}, A. 1997, \apj, 480, 188

\bibitem[{{Gould}(2004)}]{Gould04}
---. 2004, \apj, 606, 319

\bibitem[{{Gould} \& {Loeb}(1992)}]{Gould92}
{Gould}, A. \& {Loeb}, A. 1992, \apj, 396, 104

\bibitem[{{Gould} {et~al.}(2006){Gould}, {Udalski}, {An}, {Bennett}, {Zhou},
  {Dong}, {Rattenbury}, {Gaudi}, {Yock}, {Bond}, {Christie}, {Horne},
  {Anderson}, {Stanek}, {DePoy}, {Han}, {McCormick}, {Park}, {Pogge},
  {Poindexter}, {Soszy{\'n}ski}, {Szyma{\'n}ski}, {Kubiak}, {Pietrzy{\'n}ski},
  {Szewczyk}, {Wyrzykowski}, {Ulaczyk}, {Paczy{\'n}ski}, {Bramich},
  {Snodgrass}, {Steele}, {Burgdorf}, {Bode}, {Botzler}, {Mao}, \&
  {Swaving}}]{Gould06}
{Gould}, A., et al. 2006, \apjl, 644, L37

\bibitem[{{Griest} \& {Safizadeh}(1998)}]{Griest98}
{Griest}, K. \& {Safizadeh}, N. 1998, \apj, 500, 37

\bibitem[{Hardy} \& {Walker}(1995)]{Hardy95} Hardy, S.J. \& Walker, M.A. \mnras,
276, L79

\bibitem[{Holz} \& {Wald}(1996)]{Holz96}
Holz, D.E. \& Wald, R.M. 1996, \apj, 471, 64

\bibitem[{{Ida} \& {Lin}(2004)}]{Ida04}
{Ida}, S. \& {Lin}, D.~N.~C. 2004, \apj, 604, 388

\bibitem[{{Ida} \& {Lin}(2005)}]{Ida05}
---. 2005, \apj, 626, 1045

\bibitem[{{Kervella} {et~al.}(2004){Kervella}, {Th{\'e}venin}, {Di Folco}, \&
  {S{\'e}gransan}}]{Kervella04}
{Kervella}, P., {Th{\'e}venin}, F., {Di Folco}, E., \& {S{\'e}gransan}, D.
  2004, \aap, 426, 297

\bibitem[{{Laughlin} {et~al.}(2004){Laughlin}, {Bodenheimer}, \&
  {Adams}}]{Laughlin04}
{Laughlin}, G., {Bodenheimer}, P., \& {Adams}, F.~C. 2004, \apjl, 612, L73

\bibitem[{{Poindexter} {et~al.}(2005){Poindexter}, {Afonso}, {Bennett},
  {Glicenstein}, {Gould}, {Szyma{\'n}ski}, \& {Udalski}}]{Poindexter05}
{Poindexter}, S., {Afonso}, C., {Bennett}, D.~P., {Glicenstein}, J.-F.,
  {Gould}, A., {Szyma{\'n}ski}, M.~K., \& {Udalski}, A. 2005, \apj, 633, 914

\bibitem[{{Rhie} {et~al.}(2000){Rhie}, {Bennett}, {Becker}, {Peterson},
  {Fragile}, {Johnson}, {Quinn}, {Crouch}, {Gray}, {King}, {Messenger},
  {Thomson}, {Bond}, {Abe}, {Carter}, {Dodd}, {Hearnshaw}, {Honda}, {Jugaku},
  {Kabe}, {Kilmartin}, {Koribalski}, {Masuda}, {Matsubara}, {Muraki},
  {Nakamura}, {Nankivell}, {Noda}, {Rattenbury}, {Reid}, {Rumsey}, {Saito},
  {Sato}, {Sato}, {Sekiguchi}, {Sullivan}, {Sumi}, {Watase}, {Yanagisawa},
  {Yock}, \& {Yoshizawa}}]{Rhie00}
{Rhie}, S.~H., et al. 2000, \apj, 533, 378

\bibitem[{{Schechter} {et~al.}(1993){Schechter}, {Mateo}, \&
  {Saha}}]{Schechter93}
{Schechter}, P.~L., {Mateo}, M., \& {Saha}, A. 1993, \pasp, 105, 1342

\bibitem[{{Smith} {et~al.}(2003){Smith}, {Mao}, \& {Paczy{\'n}ski}}]{Smith03}
{Smith}, M.~C., {Mao}, S., \& {Paczy{\'n}ski}, B. 2003, \mnras, 339, 925

\bibitem[{{Udalski} {et~al.}(2005){Udalski}, {Jaroszy{\'n}ski},
  {Paczy{\'n}ski}, {Kubiak}, {Szyma{\'n}ski}, {Soszy{\'n}ski},
  {Pietrzy{\'n}ski}, {Ulaczyk}, {Szewczyk}, {Wyrzykowski}, {Christie}, {DePoy},
  {Dong}, {Gal-Yam}, {Gaudi}, {Gould}, {Han}, {L{\'e}pine}, {McCormick},
  {Park}, {Pogge}, {Bennett}, {Bond}, {Muraki}, {Tristram}, {Yock}, {Beaulieu},
  {Bramich}, {Dieters}, {Greenhill}, {Hill}, {Horne}, \& {Kubas}}]{Udalski05}
{Udalski}, A., et al. 2005, \apjl, 628, L109

\bibitem[{{Wozniak}(2000)}]{Wozniak00}
{Wozniak}, P.~R. 2000, Acta Astronomica, 50, 421

\bibitem[{{Yee} \& {Gaudi}(2008)}]{Yee08}
{Yee}, J.~C. \& {Gaudi}, B.~S. 2008, \apj, 688, 616

\bibitem[{{Yoo} {et~al.}(2004){Yoo}, {DePoy}, {Gal-Yam}, {Gaudi}, {Gould},
  {Han}, {Lipkin}, {Maoz}, {Ofek}, {Park}, {Pogge}, {Szyma{\'n}ski}, {Udalski},
  {Szewczyk}, {Kubiak}, {{\.Z}ebru{\'n}}, {Pietrzy{\'n}ski}, {Soszy{\'n}ski},
  \& {Wyrzykowski}}]{Yoo04}
{Yoo}, J., et al. 2004, \apj, 616, 1204

\end{thebibliography}

\end{document}